\documentclass[prl,twocolumn,showpacs]{revtex4}

\usepackage{bm}
\usepackage{graphicx}
\usepackage{times}

\newcommand*{\be}{\begin{equation}}
\newcommand*{\ee}{\end{equation}}
\newcommand{\ba}{\begin{eqnarray}}
\newcommand{\ea}{\end{eqnarray}}

\begin{document}

\title{Charge and Magnetization Inhomogeneities in Diluted Magnetic
Semiconductors}

\author{Carsten Timm}
\email{ctimm@ku.edu}
\affiliation{Institut f\"ur Theoretische Physik, Freie Universit\"at Berlin,
Arnimallee 14, D-14195 Berlin, Germany}
\affiliation{Department of Physics and Astronomy, University of Kansas,
Lawrence, KS 66045, USA}

\date{September 26, 2005}

\begin{abstract}
It is predicted that III-V diluted magnetic semiconductors can exhibit
stripe-like modulations of magnetization and carrier concentration. This
inhomogeneity results from the strong dependence of the magnetization on the
carrier concentration. Within Landau theory, a characteristic temperature
$T^\ast$ below the Curie temperature is found so that below $T^\ast$ the
equilibrium magnetization shows modulations, which are
strongly anharmonic. Wavelength and amplitude of the modulation rise for
decreasing temperature, starting from zero at $T^\ast$. Above $T^\ast$ the
equilibrium state is homogeneous, but the coupling between charge and
magnetization leads to the appearance of an electrically charged layer in
domain walls.
\end{abstract}


\maketitle

\emph{Introduction}.---Diluted magnetic semiconductors (DMS) are
investigated extensively as promising materials for spintronics applications
\cite{WAB01,ZFS04} and because of their unique physical properties
\cite{Ohn98,Tim03}. Since the magnetic interaction is mediated by
the carriers, the magnetization and the Curie
temperature increase for increasing carrier concentration
\cite{OCM00,BKB02,EWC02,NKS03,LLD05,KHT05}. In fact, the magnetization can
be changed \textit{in situ} by tuning the carrier concentration with a gate
voltage \cite{OCM00,BKB02,NKS03}.

Coupling between magnetism and carrier concentration can lead to
\emph{inhomogeneous} equilibrium states, as found
in manganites \cite{mang}, nickelates \cite{nickel}, and cuprates
\cite{copper}. The present
Letter analyzes this possibility for III-V DMS, concentrating on
\emph{stripe-like}, one-dimensional variations of the magnetizations.
We employ a Landau theory for the coupled magnetic and charge
degrees of freedom. This approach is valid on length
scales on which the impurity distribution can be treated as homogeneous
\cite{Die00}. The characteristic length scale is
$n_{\mathrm{Mn}}^{-1/3}$, where $n_{\mathrm{Mn}}$ is the density of Mn
impurities \cite{Tim03,TSO02}. Finally, we discuss possible experiments.


Magnetic domains have also been observed in (Ga,Mn)As \cite{SHF00,WVL03}.
They are formed to reduce the dipolar
energy, as in other ferromagnets. One can expect the coupling between
magnetization and carriers to lead to an inhomogeneous charge profile in a
domain wall. This question is addressed in the final part of this Letter.

\emph{Landau theory}.---We write the Hamiltonian as a functional of
magnetization $\mathbf{m}$ and deviation of carrier density from its
spatial average, $\delta n \equiv n-\overline{n}$. Charge neutrality requires
$\int d^3r\,\delta n=0$. The magnetic part has the usual form
\be
H_m = \int d^3r\, \Big\{ \frac{\alpha}{2}\,m^2
  + \frac{\beta}{4}\,m^4 + \frac{\gamma}{2}\, \partial_i\mathbf{m}
  \cdot\partial_i\mathbf{m} \Big\} ,
\label{Fm1}
\ee
where $\partial_i\equiv \partial/\partial r_i$ and
summation over $i$ is implied. The
mean-field Curie temperature is determined by $\alpha=0$. Since experimentally
the Curie temperature depends approximately linearly on carrier
concentration, we expand $\alpha=\alpha'\,(T-T_c-\eta\,\delta n)$, where
$T_c$ is the Curie temperature for $\delta n=0$. This dependence of 
$\alpha$ provides the coupling between magnetism and carrier
concentration in our model and is responsible for the physics discussed in the
following. Since the equilibrium magnetization for constant
$\alpha$ is $m_0\equiv \sqrt{-\alpha/\beta}$, larger
magnetization is favored in regions with higher carrier concentration (note
$\alpha<0$ in the ferromagnetic phase).

The second ingredient for our model is
the screened Coulomb energy due to the charge inhomogeneity,
\be
H_{\delta n} = \frac12 \int d^3r\,d^3r'\,
  \frac{e^2}{4\pi\epsilon_0\epsilon}\, \delta n(\mathbf{r})\,
  \delta n(\mathbf{r}')\, \frac{e^{-|\mathbf{r}-\mathbf{r}'|/r_0}}
  {|\mathbf{r}-\mathbf{r}'|} .
\ee
The total Hamiltonian is $H=H_m + H_{\delta n}$.

We first discuss qualitatively what kind of equilibrium states we expect from
$H$. Any inhomogeneous charge distribution \emph{increases} $H_{\delta n}$. On
the other hand, the contribution from $H_m$ is not obvious, since the first term
is negative for $T<T_c+\eta\,\delta n$. We will see that the magnetic energy
decrease in regions of higher carrier concentration and magnetization can
outweigh the increase in regions of lower $\delta n$ and $m$ and even the
increase in electrostatic energy. In that case, the equilibrium state is indeed
inhomogeneous. We consider \emph{stripe-like}, one-dimensional modulations. Two-
and three-dimensional patterns seem less likely, because they contain more
regions with large magnetization gradients for a given inhomogeneity length
scale, which increase the energy due to the gradient term in $H_m$. One could
expect the inhomogeneity to take the form of stripe domains \cite{SHF00} with
alternating magnetization. However, we will see that the equilibrium solution
shows a magnetization modulation \emph{without} sign change.



We now turn to the formal derivation of equilibrium states.
It is convenient to express $H_{\delta n}$ in terms of
the electrostatic potential $\phi$. With
$(\Delta-r_0^{-2})\phi(\mathbf{r}) = -e\,\delta
n(\mathbf{r})/\epsilon_0\epsilon$ (for p-type DMS) we
obtain the total Hamiltonian
\ba
H
& = & \int d^3r\, \bigg\{ \frac{\alpha'\,(T-T_c)}{2}\, m^2
  + \frac{\beta}{4}\, m^4 + \frac{\gamma}{2}\, \partial_i\mathbf{m}
  \cdot\partial_i\mathbf{m} \nonumber \\
& & {}+ \frac{\epsilon_0\epsilon}{2r_0^2} \phi^2
  + \frac{\epsilon_0\epsilon}{2}\, \partial_i\phi\,\partial_i\phi
  - \frac{\alpha'\eta\epsilon_0\epsilon}{2e r_0^2}\, m^2\phi \nonumber \\
& & {}- \frac{\alpha'\eta\epsilon_0\epsilon}{e}\, \mathbf{m}\cdot
  (\partial_i\mathbf{m})\partial_i\phi \bigg\} .
\label{F2}
\ea
Equilibrium configurations are given by minima of $H$
subject to the constraint of charge neutrality,
$\int d^3r\, (\Delta-r_0^{-2})\phi = 0$, which is implemented with
a Lagrange multiplier. Introducing the averaged squared magnetization
$\overline{m^2}$ and the rescaled potential $\Phi =
(\alpha'\eta\epsilon_0\epsilon/e)\, \phi$ we obtain two coupled Euler
equations for $\mathbf{m}$ and $\Phi$. Eliminating $\Phi$ from the first
we find
\ba
0 & = & -a\, \mathbf{m}\, \partial_i
  (\mathbf{m}\cdot\partial_i\mathbf{m})
  + b\,\Delta\mathbf{m} \nonumber \\
& & {}-
  \underbrace{\Big(c + \frac{a}{2r_0^2}\, \overline{m^2}\Big)}_{c'}
  \mathbf{m}
  - \underbrace{\Big(d - \frac{a}{2r_0^2} \Big)}_{d'} m^2 \mathbf{m} ,
\label{Eul1a} \\
\lefteqn{ (\Delta - r_0^{-2})\Phi \; = \; \frac{a}{2}\, (\Delta - r_0^{-2})
  (m^2-\overline{m^2}) }
\label{Eul2b}
\ea
with $a\equiv {\alpha'}^2\eta^2\epsilon_0\epsilon/e^2$, $b\equiv \gamma$,
$c\equiv \alpha'\,(T-T_c)$, $d\equiv \beta$. Equation (\ref{Eul1a})
retains information about the coupling to the carrier density since
$a\propto\eta^2$. The only bounded solution of
Eq.~(\ref{Eul2b}) is $\Phi=(a/2)\,(m^2-\overline{m^2})$. The equations
support the homogeneous mean-field solution $m^2 = m_0^2\equiv -c/d >0$,
$\Phi=0$ for $T<T_c$.
While Eqs.~(\ref{Eul1a},\ref{Eul2b}) contain five parameters,
it is sufficient to vary only \emph{two} to obtain all possible
solutions up to rescaling. We choose
$a/dr_0^2\propto \eta^2$ and $cr_0^2/b\propto T-T_c$.


\begin{figure}[t]
\includegraphics[width=3.20in,clip]{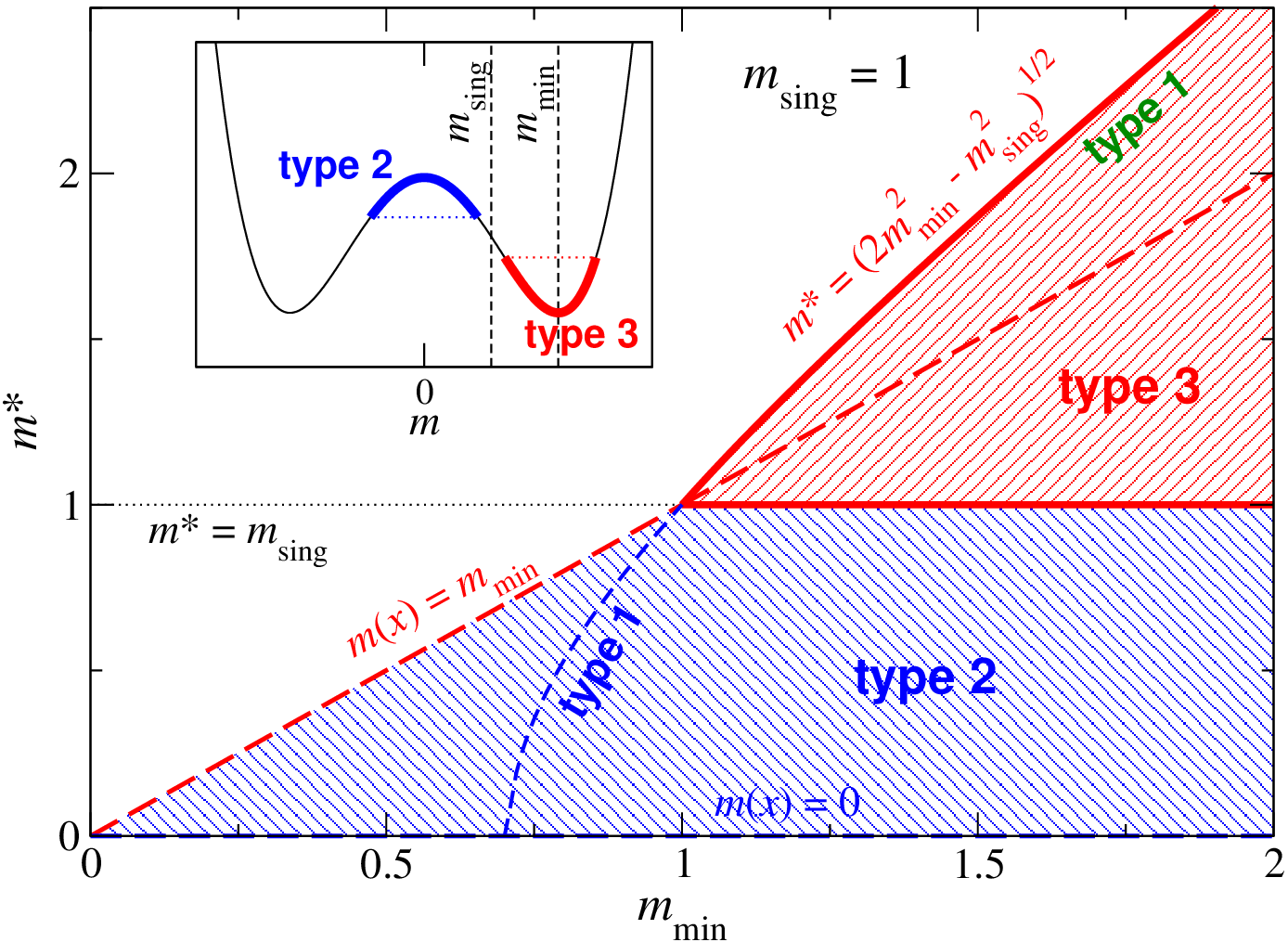}
\caption{\label{fig.pd0.ms1}(color online). Phase diagram for periodic
solutions for the magnetization in terms of $m_{\mathrm{min}}$ and
$m^\ast=m(0)$ for $m_{\mathrm{sing}}=1$. Type 1 solutions exist
for $m^\ast=(2m_{\mathrm{min}}^2-m_{\mathrm{sing}}^2)^{1/2}$ and become a
special case of type 2 for $m^\ast<m_{\mathrm{sing}}$. Type 2
solutions exist for $m^\ast<\min(m_{\mathrm{min}},m_{\mathrm{sing}})$
and type 3 solutions for
$m_{\mathrm{sing}}<m^\ast<(2m_{\mathrm{min}}^2-m_{\mathrm{sing}}^2)^{1/2}$.
The homogeneous solutions are also shown.
Inset: Schematic plot of the denominator in Eq.~(\protect\ref{invsol2})
showing the values assumed by $m$ for type 2 and 3 solutions.}
\end{figure}

\emph{Periodic solutions}.---For periodic, collinear solutions that only depend
on $x$, Eq.~(\ref{Eul1a}) becomes
\be
0 = -a\, m\, \partial_x
  (m\,\partial_x m)
  + b\,\partial_x^2 m - c' m - d' m^3 .
\label{Eul1a1}
\ee
This is an integro-differential equation due to the term
$\overline{m^2} = (1/\lambda)\int_0^\lambda dx\,m^2(x)$ in $c'$, where $\lambda$
is the wavelength. $c'$ is
determined selfconsistently below. Boundary conditions $m(0)=m^\ast$ and
$\partial_x m(0)=0$ are imposed, where $m^\ast$ will be obtained by minimizing
the energy.

Using
standard methods, we obtain the explicit integral for the inverse function
on the interval $[-\lambda/2,\lambda/2]$,
\be
x = \pm \int_{m^\ast}^m d\tilde m\,
  \sqrt{\frac{b-a\tilde m^2}{c'\tilde m^2+d' \tilde m^4/2
  -c'm^{\ast2}-d'm^{\ast4}/2}} .
\label{invsol2}
\ee
This expression satisfies $\partial_x m(0)=0$, since $\partial x/\partial m$
diverges at $m=m^\ast$. To obtain a periodic function, the denominator must
have another zero at the next extremum of $m(x)$ at $x=\pm\lambda/2$. Beyond
$\pm\lambda/2$ the solution continues periodically.

A special role is played by the magnetization value $m_{\mathrm{sing}}
\equiv \sqrt{b/a}$: Here $\partial x/\partial m$
vanishes so that coordinates $x$ beyond this point normally cannot be reached
and there is no solution for all $x$. However, a
solution (here called \emph{type 1}) crossing $m=m_{\mathrm{sing}}$ is
possible if numerator and denominator vanish \emph{simultaneously}. This
solution is $m(x) = (2m_{\mathrm{min}}^2 - m_{\mathrm{sing}}^2)^{1/2}
\cos (\sqrt{{d'}/{2a}}\,x)$ where $m_{\mathrm{min}}\equiv
\sqrt{-c'/d'}$ is the minimum of the denominator.

For all other periodic solutions $m$ must be
either larger or smaller than $m_{\mathrm{sing}}$ everywhere. For
$m^\ast<m_{\mathrm{sing}}$, periodic solutions (\emph{type 2}) oscillating
between $m^\ast$ and $-m^\ast$ with zero average exist if
$m^\ast<m_{\mathrm{min}}$. For $m^\ast>m_{\mathrm{sing}}$, periodic solutions
(\emph{type 3}) oscillating around $m_{\mathrm{min}}$ exist for
$c'm^{\ast2}+d'm^{\ast4}/2< c'm_{\mathrm{sing}}^2+d'm_{\mathrm{sing}}^4/2$,
see~Fig.~\ref{fig.pd0.ms1}.

\begin{figure}[bt]
\includegraphics[width=3.20in,clip]{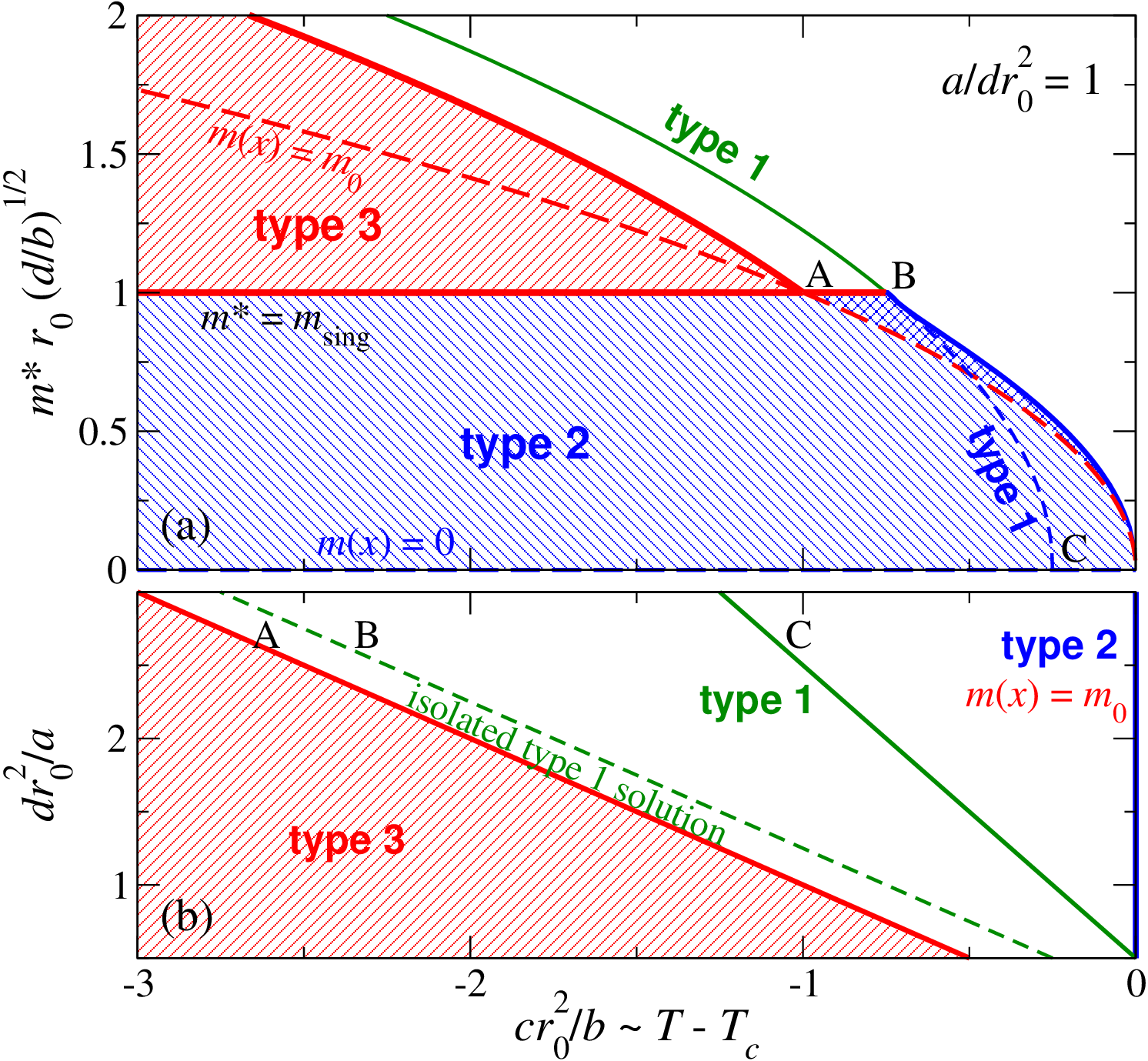}
\caption{\label{fig.pd.osc}(color online). (a) Phase diagram for periodic
solutions for the magnetization in terms of $cr_0^2/b \propto T-T_c$ and
initial magnetization $m^\ast=m(0)$ for $a/dr_0^2=1$. The
symbols are as in Fig.~\protect\ref{fig.pd0.ms1}. In the distorted triangle
with corners A,B,0, two solutions with different wavelength
coexist. (b) Phase diagram for periodic
magnetization solutions in terms of $cr_0^2/b \propto T-T_c$ and
$dr_0^2/a\propto m_{\mathrm{sing}}^2$. The various solutions exist to the
left of the respectively lines.}
\end{figure}

Next, the phase diagram is mapped back onto the parameters of the Euler equation
(\ref{Eul1a1}), determining $\overline{m^2}$ selfconsistently. The results are
shown in Fig.~\ref{fig.pd.osc}(a) for $a/dr_0^2=1$. The diagram for other values
has the same topology. Figure \ref{fig.pd.osc}(b) shows the resulting phase
diagram  in terms of $cr_0^2/b$ and \emph{general} $dr_0^2/a\propto
m_{\mathrm{sing}}^2$. The lines A, B, C show the shift of the crossing points
marked A, B, C in Fig.~\ref{fig.pd.osc}(a) with $dr_0^2/a$. Type 3 solutions
exist to the left of point A, i.e., for $m_0>m_{\mathrm{sing}}$.

From Eqs.~(\ref{F2})--(\ref{Eul2b}) we find the average energy density
\be
\overline{e} = \frac{1}{\lambda} \int_0^\lambda dx
  \left\{ -\frac{d'}{4}\, m^4
  + \frac{a}{2}\, m^2(\partial_x m)^2 \right\}
  - \frac{a}{8r_0^2}\, (\overline{m^2})^2 .
\ee
For the homogeneous solution we obtain the standard result
$\overline{e} = e_{\mathrm{hom}}\equiv
-c^2/4d$. Numerical evaluation shows that type 1 and 2 (type 3)
solutions always have higher (lower) energy than the homogeneous solution.
Among type 3 solutions the energy is minimized by the maximum
amplitude, where $m$ comes arbitrarily close to $m_{\mathrm{sing}}$.

\begin{figure}[t]
\includegraphics[width=3.00in,clip]{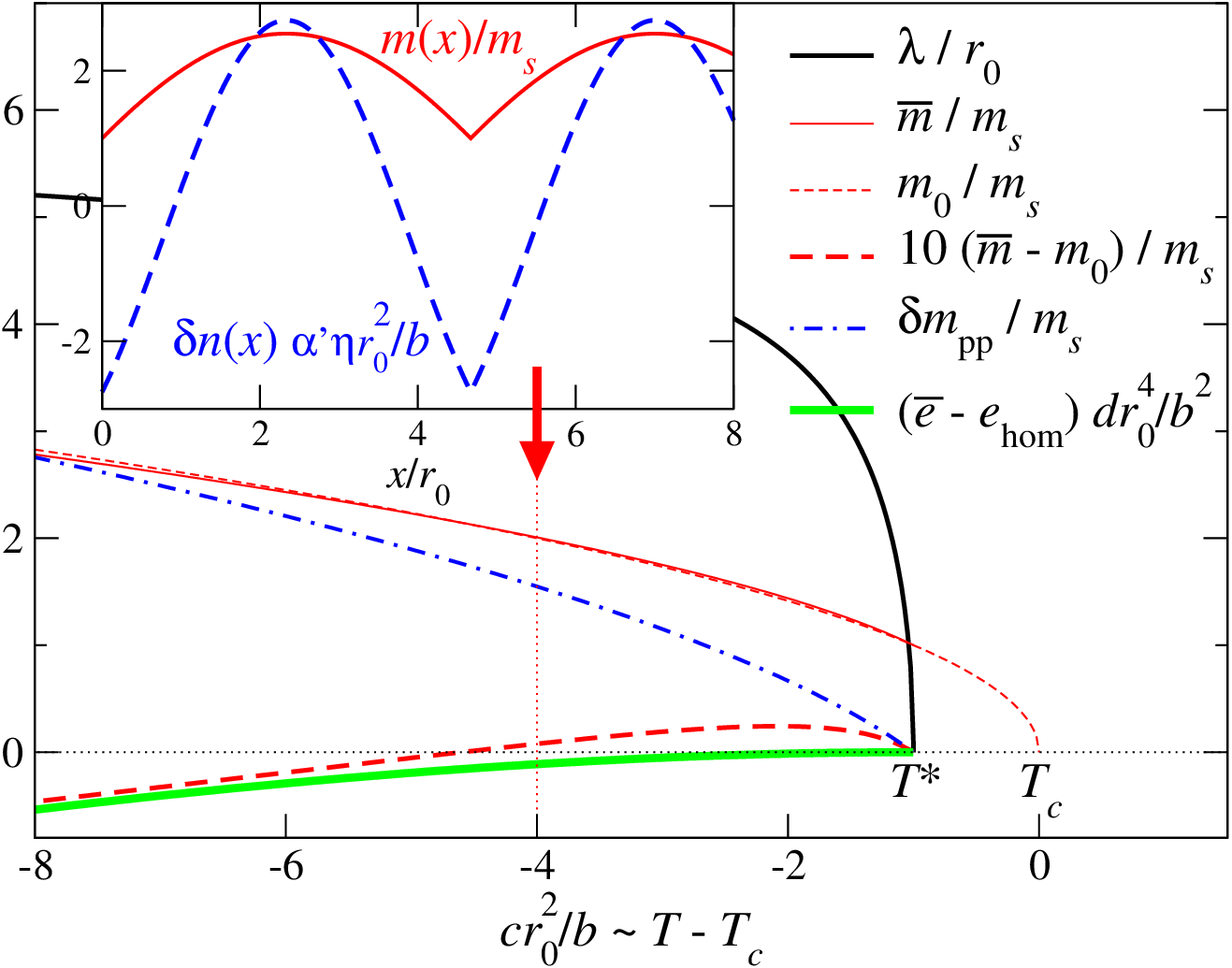}
\caption{\label{fig.optimum}(color online). Wave length $\lambda$, average
magnetization $\overline{m}$, peak-to-peak amplitude $\delta
m_{\mathrm{pp}}$, and gain in energy density $\overline{e}-e_{\mathrm{hom}}$
for the periodic magnetization with lowest energy. The magnetization for the
homogeneous solution is also shown. The unit of magnetization is $m_s\equiv
\sqrt{b/d}/r_0$. $a/dr_0^2=1$ is assumed. Inset: magnetization and
excess carrier density for $cr_0^2/b=-4$.}
\end{figure}

The mean-field magnetization of our DMS model is thus zero for $T\ge T_c$,
homogeneous for $T^\ast\le T<T_c$, where
\be
T^\ast \equiv T_c - \frac{e^2}{\epsilon_0\epsilon}\,
  \frac{\beta\gamma}{{\alpha'}^3\eta^2}
\label{Tast1}
\ee
corresponds to line A in Fig.~\ref{fig.pd.osc}(b), and a periodic
spin-density and charge-density wave for $T<T^\ast$.
The dipolar interaction omitted here
favors $\mathbf{m}$ lying in the $yz$ plane. The magnetization and potential
show sharp cusps at the minima of $m$, see the inset in
Fig.~\ref{fig.optimum}. The cusps lead to negative peaks in the
carrier density, which become $\delta$-functions for $m^\ast\to
m_{\mathrm{sing}}$. This divergence is
cut off by the condition of non-negative hole concentration. Since
the amplitude, wavelength, and energy approach finite values for $m^\ast\to
m_{\mathrm{sing}}$, the Landau theory gives a good impression of the
profile, except for some broadening of the cusps.

The optimum solution can be written down explicitly,
\be
m(x) = \sqrt{2m_{\mathrm{min}}^2-m_{\mathrm{sing}}^2} \,
  \sin\! \bigg( \sqrt{\frac{d'}{2a}}\,x + \theta \bigg)
\label{mopt2}
\ee
for $0\le x\le \lambda$ and periodically repeated.
Here, $(2m_{\mathrm{min}}^2-m_{\mathrm{sing}}^2)^{1/2}\sin\theta =
m_{\mathrm{sing}}$. From $m(x)$ one can obtain expressions for
the wavelength $\lambda = 2({2a}/{d'})^{1/2}\, ({\pi}/{2} - \theta)$,
the average magnetization $\overline{m}$, 
the peak-to-peak amplitude $\delta m_{\mathrm{pp}} =
({2m_{\mathrm{min}}^2-m_{\mathrm{sing}}^2})^{1/2} -m_{\mathrm{sing}}$, and
the energy, see Fig.~\ref{fig.optimum}.
Note that $\overline{m}$ is nonzero for all $T<T_c$.
Close to $T^\ast$ the wavelength becomes small. In this regime, the continuum
theory breaks down, since $\lambda$ is not large compared
to the disorder length scale. Figure \ref{fig.optimum} also shows that the
fundamental length scale is the screening length $r_0$. The inset
shows a typical solution.

It is important to check whether the periodic solution can occur in real
DMS. For that, $T_c-T^\ast$ should be small.
Equation (\ref{Tast1}) shows that this is the case for high
dielectric constant, small spin stiffness, strong dependence of $T_c$ on
carrier concentration, and rapid onset of magnetization below $T_c$. For
(Ga,Mn)As we estimate $T^\ast$ by comparing experiments \cite{EWC02,LLD05}
to mean-field theory for homogeneous magnetization \cite{Die00} and to
spin-wave theory \cite{JKS02}. We find $T_c-T^\ast$ of the order of
$10\,\mathrm{K}$. The properties of (Ga,Mn)As vary strongly with Mn
concentration and growth procedures. In particular,
$T_c-T^\ast$ is inversely proportional to the square of the shift of $T_c$
with carrier concentration, $\eta^2$. In Ref.~\cite{LLD05}, $\eta\sim
5.4\times 10^5\,\mbox{K\AA}{}^3$, which was used for the above estimate,
whereas in Ref.~\cite{EWC02} $\eta\sim 1.5\times 10^5\,\mbox{K\AA}{}^3$, which
would increase $T_c-T^\ast$.

Figure \ref{fig.optimum} suggests that measurements of the \emph{average}
magnetization, which have been performed extensively, are unlikely to find
evidence for the inhomogeneous state. For that, probes sensitive to the spatial
variation are required. For example, the magnetic modulation should be
observable in neutron-scattering experiments. In real space, \emph{magnetic}
scanning-tunneling microscopy (STM) and, for large $\lambda$, scanning Hall
probe experiments \cite{SHF00} or magneto-optical techniques \cite{WVL03} are
promising. Conversely, the modulation in \emph{carrier concentration} should be
observable in optical reflection or transmission for large enough $\lambda$. It
also leads to a modulation of the local density of states which could be probed
by STM. The smoking gun experiment would be to look for charge and magnetization
modulations of the same wavelength.

\emph{Domain walls}.---Finally, we study the effect of
mag\-ne\-ti\-za\-tion-car\-rier
coupling on domain walls \cite{SHF00,WVL03}. We
restrict ourselves to solutions that are homogeneous in the $y$, $z$
directions. Equations (\ref{Eul1a},\ref{Eul2b}) are solved under the
boundary conditions $\lim_{x\to\pm\infty} \mathbf{m}(x) = \pm
m_0\hat\mathbf{z}$, where $\hat\mathbf{z}$ is the unit vector in the $z$
direction. Since
$m^2(x)$ only deviates appreciably from $m_0^2$ in a finite interval, we
have $\overline{m^2}=m_0^2=-c/d$ in the limit of infinite system size,
$L\to\infty$. However, it turns out that charge neutrality can only be
satisfied by keeping terms of order $1/L$ in $\overline{m^2}$.
One such term comes from the region far from
the wall, where we write $m(L/2) \cong m_0 + m_1/L$. This means that the
enhanced carrier density compensating the reduction in the wall is spread
out over the bulk.

\begin{figure}[t]
\includegraphics[width=3.20in,clip]{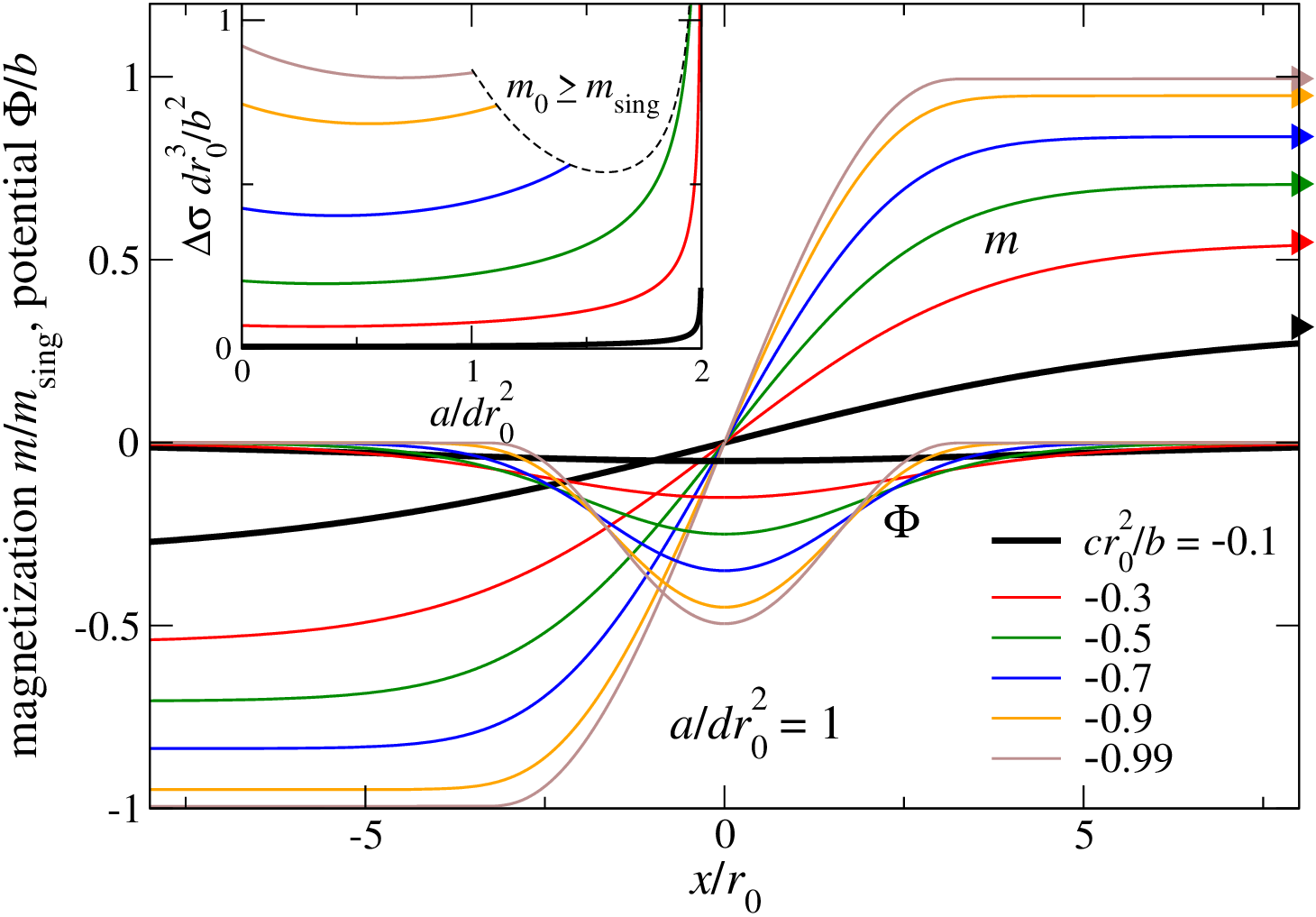}
\caption{\label{fig.typwall}(color online). Magnetization and electrostatic
potential for typical domain-wall solutions for $a/dr_0^2=1$. The
curves are for $cr_0^2/b = -0.1, -0.3, -0.5, -0.7,
-0.9, -0.99$. The asymptotical values $m_0=m(x\to\infty)$ are indicated
by triangles. Inset: Energy density $\Delta\sigma$ of a domain wall as a
function of coupling strength for the same values of $c$.}
\end{figure}

With the additional condition $m(0)=0$ we obtain
\be
x = \int_0^m d\tilde m\,
  \sqrt{\frac{b-a\tilde m^2}{c'\tilde m^2+d' \tilde m^4/2
  -c'm_0^2 - d'm_0^4/2}}
\label{xofm.wall2}
\ee
with $c'=c-(a/2r_0^2)(c/d)$ and $d'=d-a/2r_0^2$.
The integrand must be free of singularities for $0\le \tilde
m<m_0$, which implies $m_0\le m_{\mathrm{sing}}$.
Thus domain-wall solutions only exist for $T^\ast\le T<T_c$
\cite{footnote:no.wall}, see Fig.~\ref{fig.pd.osc}(a).
This indicates that $T_c-T^\ast$ is large for
samples which show domains at low temperatures.

Equation (\ref{xofm.wall2}) can be integrated explicitly.
It also yields an expression for the typical \emph{width} $\xi_w \equiv
m_0/\partial_xm(0)$ of a domain wall, $\xi_w^2 = -{2bd}/{c(d-a/2r_0^2)}$.
$\xi_w$ increases for increasing coupling $a\propto \eta^2$ between magnetism
and carriers due to their Coulomb repulsion. Figure \ref{fig.typwall} shows
$m(x)$ and $\Phi(x)$ for typical domain-wall solutions. The excess carrier
concentration $\delta n\propto (r_0^{-2}-\Delta)\Phi$ is \emph{negative} in the
domain wall, where the magnetization is reduced.

The areal energy density of the domain wall is obtained by integrating the
energy density over $x$, where corrections to $m(x)$ of order $1/L$ are again
relevant,
\ba
\Delta\sigma & = & \int_{-\infty}^\infty \!\! dx\,
  \bigg[ {-\frac{cd}{2(d+a/2r_0^2)}}\,
  \Delta m^2 - \frac{1}{4}\Big(d-\frac{a}{2r_0^2}\Big)
  \nonumber \\
& & {}\times (\Delta m^2)^2
  + \frac{a}{8}\, (\partial_x \Delta m^2)^2 \bigg]
\label{walleng2}
\ea
with $\Delta m^2\equiv m^2(\pm\infty) - m^2(x)$. The dependence of
$\Delta\sigma$ on the coupling $a\propto\eta^2$ is
shown in the inset in Fig.~\ref{fig.typwall}. $\Delta\sigma$ first decreases
with increasing coupling and then increases again, finally diverging as
$d'=d-a/2r_0^2$ goes to zero. For larger
$|cr_0^2/b|$ the divergence is not reached, since the condition
$m_0=m_{\mathrm{sing}}$ (i.e., $T=T^\ast$) is satisfied first. The initial
decrease is dominated by the $1/L$ term in $\Delta m^2(x)$ far from the
wall, i.e., from the re\-dis\-tri\-bu\-tion of carriers. In this regime,
domain walls are (slightly) less costly than they would be without
coupling. The strong increase mostly
comes from the increased width due to Coulomb repulsion. Domain walls could
be observed in the real-space experiments discussed above. The charged layer
should also affect electronic transport through domain walls.

\emph{Conclusions}.---The carrier-concentration dependence of the magnetization
in DMS introduces a characteristic temperature $T^\ast<T_c$ such that the
mean-field magnetization $m$ and excess carrier density $\delta n$ show periodic
modulations for $T<T^\ast$, whereas $m$ is homogeneous and $\delta n=0$ above
$T^\ast$. $T_c-T^\ast$ can be of the order of $10\,\mathrm{K}$ in p-type DMS.
The modulation is strongly anharmonic, and amplitude and wavelength increase for
decreasing temperature, starting from zero at $T^\ast$. For $T\ge T^\ast$ the
equilibrium state is homogeneous, but the coupling between magnetism and carrier
concentration leads to the appearance of a negatively charged layer in the
vicinity of a domain wall for p-type DMS.


\begin{thebibliography}{99}
\bibitem{footnote:no.wall}If one imposes the same boundary conditions for
$T<T^\ast$ the solution must cross the singularities at $\pm
m_{\mathrm{sing}}$. Only an oscillatory solution of type 1 is
possible.
\end{thebibliography}
\end{document}